# Dewetted Au nanoparticles on TiO$_2$ surfaces –

# Evidence of a size-independent plasmonic photo-electrochemical response

Markus Licklederer,[a] Nhat Truong Nguyen,[a] Reza Mohammadi,[b] Hyoungwon Park,[c] Seyedsina Hejazi,[a] Marcus Halik,[c] Nicolas Vogel,[b] Marco Altomare,[a]* Patrik Schmuki[a,d]*

[a] Department of Materials Science and Engineering, Institute for Surface Science and Corrosion WW4-LKO, University of Erlangen-Nuremberg, Martensstrasse 7, D-91058 Erlangen, Germany.
[b] Department of Chemical and Biological Engineering, Institute of Particle Technology, University of Erlangen-Nuremberg, Cauerstrasse 4, D-91058 Erlangen, Germany.
[c] Department of Material Science and Engineering, Institute of Polymer Materials, University of Erlangen-Nuremberg, Martensstrasse 7, D-91058 Erlangen, Germany.
[d] Chemistry Department, Faculty of Sciences, King Abdul-Aziz University, 80203 Jeddah, Saudi Arabia
* Corresponding authors. Email:  marco.altomare@fau.de
schmuki@ww.uni-erlangen.de

**Abstract**

Flat TiO$_2$ layers are deposited by magnetron sputtering on Ti/Si wafers. The TiO$_2$ surfaces are then sputter-coated with thin Au films of a nominal thickness of 0.5-10 nm that are converted by solid-state dewetting into Au nanoparticles of tuneable size and spacing; the Au nanoparticle size can be tuned over a broad range, i.e. ca. 3-200 nm. The Au-decorated TiO$_2$ surfaces enable plasmonic photo-electrochemical water splitting under visible light illumination (450-750 nm). The water splitting performance reaches a maximum for TiO$_2$ layers decorated with ~ 30 nm-sized Au particles. As expected, optical absorption measurements show a red shift of the plasmonic extinction band with increasing the Au nanoparticle size. However, the plasmonic photocurrent is found to peak at ~ 600 nm regardless of the size of the Au nanoparticles, i.e. the plasmonic photocurrent band position is size-independent. Such a remarkable observation can be ascribed to a hot electron injection cut-off effect.

**Keywords:** dewetting; Au nanoparticle; TiO$_2$; plasmon resonance; solar water splitting

## 1. Introduction

One of the most investigated metal oxide semiconductor materials for photocatalytic and photo-electrochemical generation of hydrogen is titanium dioxide ($TiO_2$). The reasons are, on the one hand, its outstanding chemical stability, nontoxicity, low cost and large availability, and, on the other hand, its semiconductive properties and electronic structure, i.e. the position of $TiO_2$ valence and conduction bands (VB and CB) is suitable to enable water splitting into $O_2$ and $H_2$ gas.[1]

Large efforts have been given in the last decades to study the photocatalytic and photo-electrochemical performance of $TiO_2$, particularly in various nanostructured forms.[1,2] Nevertheless, regardless of its morphology, a major drawback limiting the efficiency of $TiO_2$ is its "wide" band gap, i.e. of 2.8-3.2 eV depending on the crystallographic phase (i.e. anatase or rutile). The "wide" band gap sets a limit to $TiO_2$ light absorption to only UV photons, i.e. only ~ 5 % of the solar spectrum.[3]

Band gap engineering to enable visible light absorption of $TiO_2$, with e.g. non-metal elements (N-doping) has been widely explored,[4] as well as $TiO_2$ surface modification (e.g. sensitization) with dye molecules or quantum dots.[5,6,7]

Intense research activity has also focused on the possibility to combine $TiO_2$ with plasmonic metal (Au or Ag, Al) nanoparticles (NPs).[8–11] The plasmonic light absorption of metal nanostructures can potentially cover a broad visible light range upon a suitable tuning of the metal NP features (e.g. size, shape, aspect ratio, dielectric constant of the surrounding, among other parameters).[9,12–14] An advantage of using gold as the plasmonic material is its (photo)stability.[9,12–14]

Plasmon resonance can be described as the collective oscillation of the free charge in a conducting material.[15] For metal nanoparticles (NP),[10] localized surface plasmon resonance (LSPR) is stimulated by absorption of light of a suitable wavelength range. When placed on a semiconductor, plasmonic NPs can generate charge carriers, e.g. hot electrons, which can consequently be injected into the semiconductor substrate.[9,12–14] "Hot" charge carriers, generated upon visible light illumination of Au NPs supported on a TiO$_2$ photo-anode, can be used in a photo-electrochemical configuration for water splitting. For example, Au hot electrons can be injected into the semiconductor and contribute to the generated photocurrent (evolving H$_2$ gas at the Pt counter electrode), while hot holes are able to drive oxidation reactions at the Au nanoparticle surface.[16] In other words Au and TiO$_2$ can in principle act in tandem as complementary light absorbers, where TiO$_2$ and Au NPs harvest UV ($\lambda$ < 400 nm) and visible (400 nm < $\lambda$ < 800 nm) light, respectively.

Plasmon-induced photo-electrochemical activity has been observed for different types of Au-TiO$_2$ photo-electrodes, where TiO$_2$ can be in various geometries, e.g. ranging from a "planar architecture" such as compact or mesoporous (nanoparticulate) films, or in the form of arrays of one-dimensional (1D) nanostructures e.g. nanotubes, nanowires and nanorods. Plasmonic Au nanoparticles can, in a simplest approach, be formed by wet chemical synthesis from Au precursors, using reducing chemical agents, or by photo-deposition.[9,11,16–21]

In the present work we produce Au nanoparticles by solid state dewetting of thin Au films deposited by Ar plasma sputtering on flat TiO$_2$ layers. The TiO$_2$ layers are fabricated by reactive magnetron sputtering of Ti in an Ar-O$_2$ plasma, and feature a surface roughness < 1 nm. The flat and smooth surface of the TiO$_2$ substrates allows for dewetting of Au films to occur in an ordered fashion, thus forming defined Au particles of tunable size and distribution; Au nanoparticle size and spacing can be finely controlled by the initial thickness of the Au films.

We find that Au nanoparticle dewetted at TiO$_2$ rutile surfaces enable solar photo-electrochemical water splitting, and enable the generation of plasmonic photocurrent in the 450-750 nm range. The photocurrent is maximized when the dewetted Au nanoparticles are ~ 30 nm-sized.

Interestingly, while optical absorption measurements show a red shift of the plasmonic extinction band with increasing the Au NP size, we observe that the plasmonic photocurrent peaks at ~ 600 nm regardless of the Au particle size. Such a discrepancy between the optical absorbance and the photo-electrochemical features can be ascribed to a hot electron injection cut off effect, namely due to different size-dependent plasmon decay mechanisms.

## 2. Experimental Section

The $TiO_2$ surfaces were produced by magnetron sputtering in an ultra-high vacuum (UHV) chamber (Createc, SP-P-US-6M-3Z) on Ti metal-coated Si wafer (3 inch, Microchemicals, Si (100) p-type). Control experiments were carried out by producing the $TiO_2$ surfaces also on other Ti-coated substrates, such as 3 inch $SiO_2$/Si wafer (Microchemicals, Si (100) p-type + 100nm $SiO_2$) and quartz glass slides (GBV, Germany).

The titanium metal coating was sputtered (e.g. onto the Si substrate) from a Ti target (Hauner HMW, 99.999 %) operated in direct current (DC) mode, with a deposition power of 150 W, at room temperature. During deposition, the working pressure was held constant at 1 x $10^{-3}$ mbar and the Ti metal deposition time was 10 min, forming a Ti metal film with a thickness of 75 nm. Afterwards the Ti metal-coated Si substrates were patterned with a frame mask, leaving behind free (unmasked) Ti patches with a size of 1.4 x 2.4 $cm^2$. The unmasked Ti surface was then coated with amorphous 75 nm-thick $TiO_2$ layers deposited by reactive magnetron sputtering, with a deposition power of 150 W, at room temperature, in an Ar-$O_2$ plasma (Ar-$O_2$ ratio 1:5). During deposition, the working pressure was held constant at 1 x $10^{-3}$ mbar. The deposition time was 60 min, forming a $TiO_2$ film with a thickness of 75 nm.

The $TiO_2$ surfaces formed on Ti/Si substrates were then cut into 2.5 x 3.0 $cm^2$ samples, having a Ti-coated frame (surrounding the $TiO_2$-coated patch) that served as back-side electric contact (electron collector scaffold) in the photo-electrochemical measurements.

The samples were then coated with Au films of different thicknesses (0.5-10 nm) using a plasma sputtering device (EM SCD500, Leica). The amount of Au, in terms of nominal Au film thickness, was controlled by an automated quartz crystal microbalance (thickness monitor).

The Au-coated $TiO_2$ substrates were then annealed in a tubular furnace at 450°C in air for 1h. This thermal treatment is key as it simultaneously leads to (i) crystallization of the amorphous $TiO_2$ surface into rutile phase, and (ii) solid-state dewetting of the Au films into Au nanoparticles of controllable size and spacing.

For the optical measurements (experimental described below), Au NPs were formed by argon plasma deposition of Au films of different thicknesses (0.5-10 nm) on quartz glass slides (1.5 x 1.5 $cm^2$, GVB, Germany). The quartz substrates were previously cleaned by sonication in acetone, ethanol and deionized water (15 min for each step), and then dried in a $N_2$ stream. After the deposition, the quartz supported Au films were dewetted in a tubular furnace at 450°C in air (1 h).

The morphology of the Au-decorated rutile $TiO_2$ surfaces was characterized using a field emission scanning electron microscope (Hitachi FE-SEM, 4800). The chemical composition of the samples was carried out by X-ray photoelectron spectroscopy (XPS, PHI 5600, US). XPS spectra were acquired using Al X-ray source. The XPS spectra were corrected in relation to the C 1s signal at 284.6 eV. PHI MultiPak™ software and database were used for quantitative analysis. X-ray diffraction (XRD) with an X′pert Philips MPD (equipped with a Panalytical X'celerator detector) was used to examine the crystallographic properties of the materials. Depth profiling was carried out using the instrument's $Ar^+$ sputter source operated at 3 kV, 15 nA, rastered over a 3 × 3 $mm^2$ area with a sputtering angle of 45° to the surface normal. 1 min long sputter steps were repeated to analyze in depth the composition of the whole architectures until reaching the Si substrate. The atomic composition was determined between consecutive sputtering intervals, by evaluation the photoelectron peak areas, using the MultiPak™ processing software.

AFM imaging was performed using a Veeco Dimension 3100 Microscope, operated in tapping mode. Bruker silicon probes with a spring constant of 2 N m$^{-1}$ and 70 kHz resonance frequency were used. The resolution was set to 1024 × 1024 pixels for measured surface of 2 × 2 or 10 × 10 µm². Processed data were acquired with Gwyddion 2.42. The RMS roughness values were evaluated by the statistical quantities tool embedded in the Gwyddion SPM analysis software. The lower threshold was set to 0.05 nm to correct for sample roughness.

The IPCE spectra were recorded with a setup comprising an Oriel 6356 150 W Xe arc lamp as light source and an Oriel cornerstone 7400 1/8 monochromator. The measurements were carried out in aqueous 0.1 M $Na_2SO_4$ solutions at an applied potential of 0.5 V in an electrochemical cell equipped with a quartz glass window and operated in a three electrode configuration with the Au-$TiO_2$ structures used as working electrode, a Ag/AgCl electrode as reference and a Pt foil as counter electrode. Photo-current transients under monochromatic light illumination (600 nm) were also recorded under constant external applied bias (+500 mV) and 20s-10s light on-off cycles.

The photo-electrochemical water-splitting performance of the samples was investigated in aqueous 1 M KOH solutions with a three-electrode configuration: the Au-$TiO_2$ structures were used as working electrode, a saturated Ag/AgCl electrode was used as the reference and a platinum foil was used as the counter electrode. For the photo-electrochemical linear sweep voltammetry (LSV) experiments, an external potential was applied to the photoelectrochemical cell provided by a scanning potentiostat (Jaissle IMP 88 PC), and was swept from -0.65 V to +1.25 V. The experiments were carried out under simulated AM 1.5 illumination provided by a solar simulator (300 W Xe with a Solarlight optical filter). The light intensity emitted by the solar simulator (operated without the cut off filter) was measured prior to the experiments using a calibrated Si photodiode and the distance between the light source and the photo-anode was adjusted to illuminate the samples with an irradiation power of 100 mW cm$^2$. Photo-current transients under visible light illumination (solar simulator with 420 nm cut-off filter) and under illumination with a 520 nm laser (1 W, OdicForce Lasers, OFC 420-G1000) were also recorded under a constant applied bias (+500 mV) and under 30s-10s light on-off cycles.

To study the optical properties of the quartz supported Au nanoparticles, we used a UV-Vis-IR spectrometer (Perkin-Elmer Lamda950) with an integrating sphere and illumination beam of approximately 8 mm in diameter. We performed the transmission measurements from 300 to 1500 nm with 2 nm resolution.

## 3. Results and Discussion

### 3.1 Physicochemical characterization

The structure of the plasmonic Au-$TiO_2$ surfaces produced in this work is outlined in Fig. 1a. The $TiO_2$ surfaces are produced on Ti-coated Si wafer substrates by magnetron sputtering. The SEM analysis of a cross-sectional cut of a typical structure is shown in Fig. 1b, where one can see (from bottom to top) the Si/Ti and Ti/$TiO_2$ interfaces (highlighted with dotted lines). The $TiO_2$ and Ti layers both have a thickness of 75 nm.

The XRD data in Fig. 1c reveal that in the as-deposited architecture, the $TiO_2$ surface is amorphous; no reflections of crystalline $TiO_2$ polymorphs can be seen. However, a thermal treatment in air at 450°C (1 h) leads to crystallization into rutile phase. This is evident from the appearance of a peak at 27.5° upon annealing, which can be attributed to the $TiO_2$ rutile (110) reflection.[22] Additional XRD data (measured in the 20-80° 2θ range) confirm the formation of $TiO_2$ rutile phase – provided in Fig. S1a. The XRD data in Fig. S1b were collected by θ-2θ rocking curve scans. Aside from the Si peak at 70° (ascribed to the single crystal Si substrate) one can notice that even upon annealing the Ti bottom layer remains in the metallic state; this is proved by the presence of a Ti reflecction at 38°.

The crystalline $TiO_2$ surfaces were characterized by XPS in view of their chemical composition. The XPS survey in Fig. S1c shows the layers to be composed of Ti and O (and adventitious C).

We found a surface 1:2 Ti:O compositional ratio (Ti ~ 33% and O ~ 67%), which is in line with the $TiO_2$ stoichiometry. The high resolution spectra in the Ti 2p and O 1s regions (Fig. S1d,e) show peaks that fit well to $TiO_2$ XPS data reported in the literature.[23] Such Ti 2p and O 1s peaks seem not to vary significantly upon annealing and conversion of the amorphous $TiO_2$ into crystalline rutile phase.

The XPS sputter profile of a typical $TiO_2$/Ti layer is presented in Fig. 1d (both before and after $TiO_2$ crystallization). In both cases, one can identify two interfaces, i.e. the $TiO_2$/Ti and Ti/Si interfaces (from top to bottom), at a depth of approximately 75 and 150 nm, respectively.

The surface roughness of as-deposited and crystallized $TiO_2$ surfaces was investigated by AFM measurements (Fig. 1e and Fig. 1f, respectively). As-deposited $TiO_2$ surfaces show a RMS surface roughness < 1 nm (precisely, 9.6 Å). Upon annealing, the $TiO_2$ surface roughness is found to increase to 4.5 nm; this is ascribed to $TiO_2$ crystallization into rutile phase, and consequent grain growth and faceting. However, the crystalline $TiO_2$ surfaces remain homogeneous and conformally-coated on the Ti/Si substrates, and no through-layer pinholes could be observed.

The Au nanoparticles are formed on the $TiO_2$ surfaces by solid-state "dewetting" of sputtered Au films.[24,25] Firstly, Au thin films with a nominal thickness in the 0.5-10 nm range are deposited on the amorphous $TiO_2$ surfaces. The Au films were then converted into self-ordered Au particles by thermal dewetting. Dewetting occurs as thin metal films are unstable in the as-deposited state and when heated up to a certain temperature, tend to split-open and agglomerate, forming metal particles via surface diffusion.[26] In our recent work, we explored metal dewetting phenomena at the surface of semiconductor metal oxides, e.g. $TiO_2$, to fabricate photocatalysts and photo-electrodes.[26–29]

In the present work, the Au-films deposited at the $TiO_2$ surfaces were dewetted by a thermal treatment in air at 450°C (1 h); this is well in line with the melting point of Au ($T_{m,Au}$ = 1064°C), as depending on the metal film thickness, solid state dewetting is expected to occur at temperatures well below the metal melting point, e.g. between 1/3 and 1/2 of $T_{m,Au}$.[24,25] Dewetting occurs (vide supra) together with the crystallization of the amorphous oxide surface into rutile $TiO_2$ phase.

The pristine and Au-decorated $TiO_2$ surfaces were analyzed by SEM. The results are compiled in Fig. 2 (additional SEM data are in Fig. S2). The SEM images in Fig. 2b-g (and Fig. S2) demonstrate that the size (and spacing) of the dewetted Au NPs can be well controlled over a broad range (from ca. 3 to 200 nm). The data in Fig. S3a show the size distribution statistics of Au NPs dewetted from Au films of different initial nominal thicknesses. Table S1 and Fig. S3b summarize the results. We observe, in line with previous work, that the thicker the Au film, the larger the Au NP size and spacing, and the broader their size distribution statistics.[24,25] For example, dewetting of a 0.5 nm-thick Au film forms 3.2 nm-sized Au NPs; a 2 nm-thick Au film forms instead 32 nm-sized Au particles (i.e. one order of magnitude bigger NPs). Also, it seems that within the size range explored in this work, the Au NP size scales linearly with the Au film initial thickness (Fig. S3b).

*3.2 Photo-electrochemical characterization*

The pristine and Au-decorated $TiO_2$ surfaces were investigated as photo-anodes for solar photo-electrochemical water splitting. The photo-electrochemical results are compiled in Fig. 3. Fig. 3a shows the IPCE spectra of pristine and $TiO_2$ surfaces decorated with dewetted Au NPs of different sizes. The Au-$TiO_2$ samples are labeled according to the initial nominal thickness of the sputtered Au film.

Pristine $TiO_2$ surfaces (sample "0 nm") generate photocurrent only upon band gap excitation, i.e. under UV light illumination. Well in line with the rutile phase composition, the photocurrent onset is at ~ 410 nm, corresponding to a band gap energy of 3.0 eV (see band gap estimation in Fig. S4).[30] On contrary, all the Au-decorated $TiO_2$ surfaces generate photocurrent not only upon

TiO$_2$ bang gap excitation, but also under visible light illumination, showing an evident photocurrent peak in the 450-750 nm range (inset in Fig. 3a).

In the following sections we examine and discuss the different contributions to the photocurrent, i.e. ascribed to TiO$_2$ UV light absorbance or Au plasmon resonance, in relation to parameters such as Au NP size, free TiO$_2$ surface, shading effects, hot carrier generation and extraction efficiency, and electron trapping effects.

*3.3 UV light-induced photocurrent*

In the UV region, the highest IPCE values are recorded for the pristine TiO$_2$ rutile surfaces (0 nm), which deliver a maximum IPCE of 20 % measured at 300 nm. Overall, the photocurrent trend at 300 nm is 0 nm > 2 nm > 5 nm > 3 nm > 10 nm > 1 nm >> 0.5 nm. To interpret these results one may take into account that the photocurrent generation under UV light illumination may be affected by factors such as (i) light shading effects ascribed to the Au NPs decorated at the TiO$_2$ surface – this effect is expected to set in mainly for large Au particles (e.g. > 50 nm); and (ii) the extent of free (uncoated) TiO$_2$ surface. The former factor (Au NP size and surface coverage) can influence the photon flux reaching the TiO$_2$ surface. The latter is connected to the efficiency of TiO$_2$ hole injection into the electrolyte; it is reasonable to assume that the ease of hole transfer to the electrolyte increases with increasing the free (uncoated) TiO$_2$ surface, while hole transfer becomes gradually more hampered with increasing the Au surface coverage (the consequent hole accumulation in TiO$_2$ can result in charge recombination and thus in a limited photocurrent). When plotting the free TiO$_2$ surfaces and the IPCE (measured at 300 nm) as a function of the Au film initial thickness (Fig S5), no straightforward correlation between these parameters can be found.

Hence, we propose that the photocurrent measured under UV light illumination may be also affected by the ability of Au NPs to trap TiO$_2$ conduction band electrons. In a PEC configuration under the sole UV light illumination, the photocurrent can only be caused by photo-promoted electrons in TiO$_2$ (band gap excitation). These electrons have to diffuse through the TiO$_2$ rutile layer to be collected by the (Ti metal) back contact. In spite of the applied anodic bias (that causes upward band bending in TiO$_2$ and thus generate an electron-depleted space charge region in the semiconductor) it may however occur that photo-promoted electrons (more likely those generated at the surface or sub-surface of the TiO$_2$ photo-anode) can be trapped by Au NPs. In previous work, the deposition of noble metal NPs on the surface of TiO$_2$ photo-anodes was observed to lead to an evident UV photocurrent drop,[31–33] although such nanoparticles had an average size < 10 nm and were thus not expected to contribute to UV light shading effects. Interestingly, Fig. S6 shows the trend of the Au NP surface decoration density and IPCE (measured at 300 nm) as a function of the Au film initial thickness. The lowest IPCE values are measured for the highest Au NP decoration density (which are obtained by dewetting 0.5 nm and 1 nm-thick Au layers). These results seem to point to a trapping effect for UV light-generated TiO$_2$ CB electrons that becomes dominant for high surface decoration densities of Au-TiO$_2$ Schottky junctions, i.e. the higher the decoration density (number of Au NPs per unit area), the larger the electron trapping effect and, consequently, the lower the UV photocurrent.

*3.4 Plasmonic photocurrent*

The photocurrent observed in the visible light region (inset in Fig. 3a) can be ascribed to surface plasmon resonance (SPR) effects enabled by the Au NPs dewetted at the TiO$_2$ surfaces. Please note that no visible light photocurrent can be observed for pristine TiO$_2$ surfaces. The photocurrent results are summarized in Fig. 3b as a function of the Au film initial thickness. Fig. 3c illustrates the photocurrent transients measured for the different photo-anodes under monochromatic visible light (600 nm) illumination. From Fig. 3b one can see that the photocurrent response increases with increasing the Au film thickness from 0.5 nm to 2 nm. A further increase of the Au film thickness is found detrimental as it leads to a remarkable photocurrent drop. A Au film with an initial thickness of 2 nm, which forms by dewetting ~ 30 nm-sized Au NPs at the TiO$_2$ surface, leads to the highest plasmonic photocurrent response, corresponding to an IPCE of 0.4 % (inset in Fig. 3a).

Previous work on plasmonic metal nanoparticles on a semiconductor surface demonstrated that the LSPR effect, in terms of visible light absorption cross section, hot carrier generation efficiency and intensity of the plasmon-induced electromagnetic field, is dependent on the metal particle size and spacing.[9] For Au NPs on TiO$_2$, the absorption band associated to plasmon resonance falls at a specific wavelength in the 500-600 nm range as a function of the Au NP shape and size.[34–38] Our results, however, show that the plasmonic effect varies as a function of the Au NP size only in terms of photocurrent magnitude (photocurrent density), while the band position is found to be constant at ~ 600 nm regardless of the Au NP size.

To clarify this aspect, we measured the optical absorbance of Au NPs dewetted on quartz glass slides (data compiled in Fig. S7a,b). In average, the extinction maxima are observed in the 520-570 nm region, i.e. are blue shifted compared to the plasmonic photocurrent peaks; this is due to the refractive index of the substrate (quartz glass vs. TiO$_2$). In other words, the Au plasmon band measured by IPCE at longer wavelength ($\lambda_{max}$ ~ 600 nm) is due to the higher refractivity index of TiO$_2$ compared to quartz.[39] More importantly, the optical measurements show a red shift of the plasmonic extinction band maxima ($\Delta\lambda$ = 14 nm) with increasing the Au NP size (initial Au film thickness increasing from 0.5 nm to 5 nm); see data in Fig. S7c. The red shift is accompanied by an evident peak broadening. Particles dewetted from a 10 nm thick Au film show an even more pronounced red shift ($\Delta\lambda$ = 58 nm) and a dramatic band widening effect (see the SI for more details).

Such a mismatch between the optical absorbance and PEC performance, and the size-independence of the plasmonic photocurrent peak, can be ascribed to a hot electron injection cut off effect. In a PEC configuration, plasmonic electrons can contribute to the photocurrent only if they are injected into the CB of the semiconductor substrate. Our optical measurements show that the larger the Au particles the longer the wavelength of maximum plasmonic extinction. Interestingly, previous spectroscopic investigations demonstrated that plasmonic decay substantially varies in relation to the excitation wavelength, while it appears almost unaffected by the Au particle size.[40] More precisely, long-lived hot carriers were found to be generated by absorption of photons of relatively lower wavelength (e.g. < ~ 560 nm). On the contrary, hot electrons excited by longer wavelength were found to undergo rapid decay; in other words, bigger Au particles do absorb longer wavelength photons but the generated hot electrons recombine rapidly and thus cannot effectively contribute to the photocurrent; on the other hand, the relatively long lifetime of hot electrons excited with shorter wavelength photons allows for a more efficient charge extraction and injection into the $TiO_2$ surfaces. In our view, aside from a slower plasmon decay mechanism, hot electrons excited with shorter wavelength photons may be more efficiently injected into the semiconductor CB also owing to their higher energy level, which allows them to overcome the Schottky barrier at the $Au/TiO_2$ interface. In the case of our results, we can assume that such a hot electron injection energy cut off effect has a threshold wavelength at ~ 600 nm. At ~ 600 nm, hot electron excitation efficiently contributes to photocurrent. Below this wavelength the plasmonic extinction is limited (low optical absorption and hence small density of hot charge carriers), while above 600 nm the hot electrons may not possess sufficient energy to overcome the Schottky barrier. At the threshold wavelength (600 nm), the photocurrent is mainly determined by the Au NP size: a maximum charge separation and injection efficiency is observed for ~ 30 nm-sized Au NPs, as also reported in literature.[41,42] In contrast, Au NPs < 10 nm only enable a relatively low light absorption, due to surface damping effects (e.g. Landau damping), while Au NPs > 40 nm typically lead to a less efficient hot electron injection into $TiO_2$ (it is in fact reported that the injection efficiency decreases with increasing the Au particle volume, keeping the particle aspect ratio constant).[41,42]

For $Au-TiO_2$ structures formed from Au films with thickness in the 1-5 nm range, the photocurrent transients resemble a square wave signal, i.e. the response is characterized by an instantaneous photocurrent increase up to a maximum photocurrent value upon light illumination, followed by a rapid photocurrent drop in the dark (Fig. 3c). In contrast, the response is relatively slow for sample 10 nm, and particularly sluggish (and "noisy") for sample 0.5 nm; in line with what above, the origin of the sluggish photocurrent response for large Au NPs (dewetted from 10 nm thick Au film) can be a poor injection efficiency, while the reason for the sluggish and noisy response observed for sample 0.5 nm has yet to be clarified.

*3.5 Solar photocurrent*

The solar PEC water splitting ability of the photo-anodes was investigated by photo-electrochemical linear sweep voltammetry (LSV) experiments. Measurements were also performed by applying to an AM 1.5 light source (solar simulator) a 420 nm cutoff filter, in order to illuminate the photo-anodes with visible light only. The results (Fig. 3d) show that the $I_{ph}$ onset is in any case at -500 mV. With the 420 nm cutoff filter, the Au NP-decorated $TiO_2$ surfaces generate photocurrent, while no photocurrent is observed for the reference pristine $TiO_2$.

The PEC results under full lamp illumination show that the water splitting ability of the Au-$TiO_2$ photo-anodes is higher than that of pristine counterparts; the Au plasmon resonance increases the photoresponse from 10.8 $\mu A\ cm^{-2}$ to 12.6 $\mu A\ cm^{-2}$ (at +500 mV vs. Ag/AgCl), which corresponds to a photocurrent increase of 16%. Please note that under the sole UV light illumination, the photocurrent of Au-$TiO_2$ structures was found to be lower than that of pristine $TiO_2$ layers (Fig. 3a). Therefore, one can conclude that under full lamp illumination (simulated solar light) the plasmonic enhancement of the Au NPs counterbalances the loss of photocurrent in the UV light range (ascribed to trapping effects; see section 3.3).

Interestingly, the net photocurrent difference ($\Delta I_{ph,\ +1000mV}$) between Au-$TiO_2$ and pristine $TiO_2$ photo-anodes measured under full lamp illumination is 3.6 $\mu A\ cm^{-2}$, which is remarkably smaller than the photocurrent measured for Au-$TiO_2$ photo-anodes with the 420 nm cutoff filter, i.e. 8.0 µA cm$^{-2}$. This is also evident when comparing data in Fig. 3d with the photocurrent transients in Fig. S8a – measured at +500 mV vs. Ag/AgCl under AM 1.5 illumination with 420 nm cutoff filter. In addition, the photocurrent transients in Fig. S8b, measured under chopped illumination provided by a 520 nm laser, further validate the plasmonic contribution and the stability of the plasmonic effect. More importantly, the reason for such a discrepancy can be that under combined UV and visible light illumination, the extraction of hot electrons from Au can be less favorable as the concurrent promotion of TiO$_2$ CB electrons under UV light illumination leads to an upward shift of the semiconductor Fermi level, which in turn increases the height of the Au-TiO$_2$ Schottky junction, thus reducing to some extent the hot electron injection efficiency. The visible light photocurrent ($\lambda > 420$ nm) generated by the Au-TiO$_2$ structure is 8.0 µA cm$^{-2}$ at +1000 mV, which is more than 10 times higher than that reported in the literature for comparable architectures,[43] and is even 8 times higher than that of thicker TiO$_2$ photo-anodes tested under higher applied potentials.[16]

Finally, to rule out the possibility that the observed visible light photocurrent my originate from artefacts, e.g. from the Si substrates, we produced TiO$_2$ surfaces identical to those described above, both pristine and decorated with dewetted Au NPs, on other (non-conductive) substrates such as Ti metal-coated SiO$_2$/Si wafers and quartz glass slides (see Scheme S1). The Au-decoration of these architectures was obtained by dewetting (450°C, 1 h, air) Au films with an initial thickness of 2 nm. The structures were tested as photo-anodes for PEC water splitting. The results (Fig. 3e,f) show that the Au-TiO$_2$ architectures generate visible light photocurrent even when formed on Ti/SiO$_2$/Si wafers and Ti/quartz glass slides.

## 4. Conclusions

In this work we fabricated Au nanoparticle-decorated TiO$_2$ surfaces for photo-electrochemical water splitting. Au nanoparticles were formed by solid state dewetting from sputtered thin Au films. Size and spacing of the nanoparticles could be controlled by adjusting the Au film initial thickness. As a result, a fine tune of the Au NP size in the ~ 3-200 nm range was enabled. The photo-electrochemical performance and optical features of the Au-TiO$_2$ surfaces were investigated in relation to the Au NP size and decoration density, considering parameters such as free TiO$_2$ surface, shading effects, hot carrier generation and extraction efficiency, and electron trapping effects. All Au-decorated photo-anodes enabled solar photo-electrochemical water splitting, showing a plasmonic photocurrent in the visible range (450-750 nm). The photocurrent response was the highest for ~ 30 nm-sized Au nanoparticles. As expected, optical absorption measurements show a red shift of the plasmonic extinction with increasing the Au NP size. However, the plasmonic photocurrent peaked at ~ 600 nm regardless of the Au particle size. Such a discrepancy is caused by a hot electron injection cut off effect that sets-in in a photo-electrochemical configuration due to the wavelength-dependence nature of the plasmonic decay. Moreover, the Au-TiO$_2$ surfaces showed a 16% higher solar PEC water splitting performance compared to pristine TiO$_2$ surfaces, and the plasmonic response measured under the sole visible light illumination was found to be significantly higher than that achieved under simultaneous with UV and visible light illumination. We propose that the UV semiconductor activation (band gap excitation) can limit the plasmonic hot electron injection from Au nanoparticles into TiO$_2$ due to an upward shift of the semiconductor Fermi level which in turns partially hinders the hot electron injection efficiency, owing to an increase of the Schottky barrier height.

**Acknowledgements**

The authors acknowledge the ERC, DFG, and DFG Cluster of Excellence EAM for financial support. The authors would like to acknowledge Ulrike Martens for the XRD analysis, Anja Friedrich for the SEM investigation, and Helga Hildebrand for the XPS measurements. Dr. Johannes Will (Chair of Micro- and Nanostructure Research, University of Erlangen-Nuremberg, Erlangen, Germany) is also acknowledged for technical help with the XRD measurements.

**Figure captions**

**Figure 1:** A) Schematic and B-F) physicochemical characterization of $TiO_2$ surfaces (formed on Ti/Si wafers) decorated with dewetted Au nanoparticles. B) Cross-sectional SEM image of the sample after annealing (450°C, 1 h in air). C) X-ray diffraction pattern of as-sputtered and annealed (450°C, air, 1 h) architectures. D) XPS depth profile measurement of as-sputtered and annealed architectures. E) and F) AFM analysis of as-sputtered and annealed architectures (root mean squared (RMS) surface roughness results are reported as inset in E) and F)).

**Figure 2:** Top view SEM images of $TiO_2$ surfaces (formed on Ti/Si wafers) sputter-coated with Au films of various thicknesses: A) 0 nm, B) 0.5 nm, C) 1 nm, D) 2 nm, E) 3 nm, F) 5 nm, and G) 10 nm; all the structures were subjected to a thermal treatment in air at 450°C (1 h) to induce dewetting of the Au films into Au nanoparticles.

**Figure 3**: Photo-electrochemical (PEC) characterization of different Au NP-decorated $TiO_2$ surfaces (formed on Ti/Si wafers). A) IPCE spectra of $TiO_2$ surfaces coated with Au films of different thicknesses and then dewetted. B) Photocurrent density and IPCE values measured at 600 nm as a function of the Au film initial thickness. C) Photocurrent transients measured at 600 nm. D) Photoelectrochemical water-splitting behavior (linear sweep voltammetry) for pristine and Au NP-decorated $TiO_2$ surfaces measured under full lamp illumination and with 420 nm cut off filter. E) IPCE spectra of different pristine and Au NP-decorated $TiO_2$ surfaces formed on Ti/Si wafers, Ti/$SiO_2$/Si wafers and Ti-coated quartz slides. F) Magnified view (visible light range) of the IPCE spectra in E).

**Figure 1**

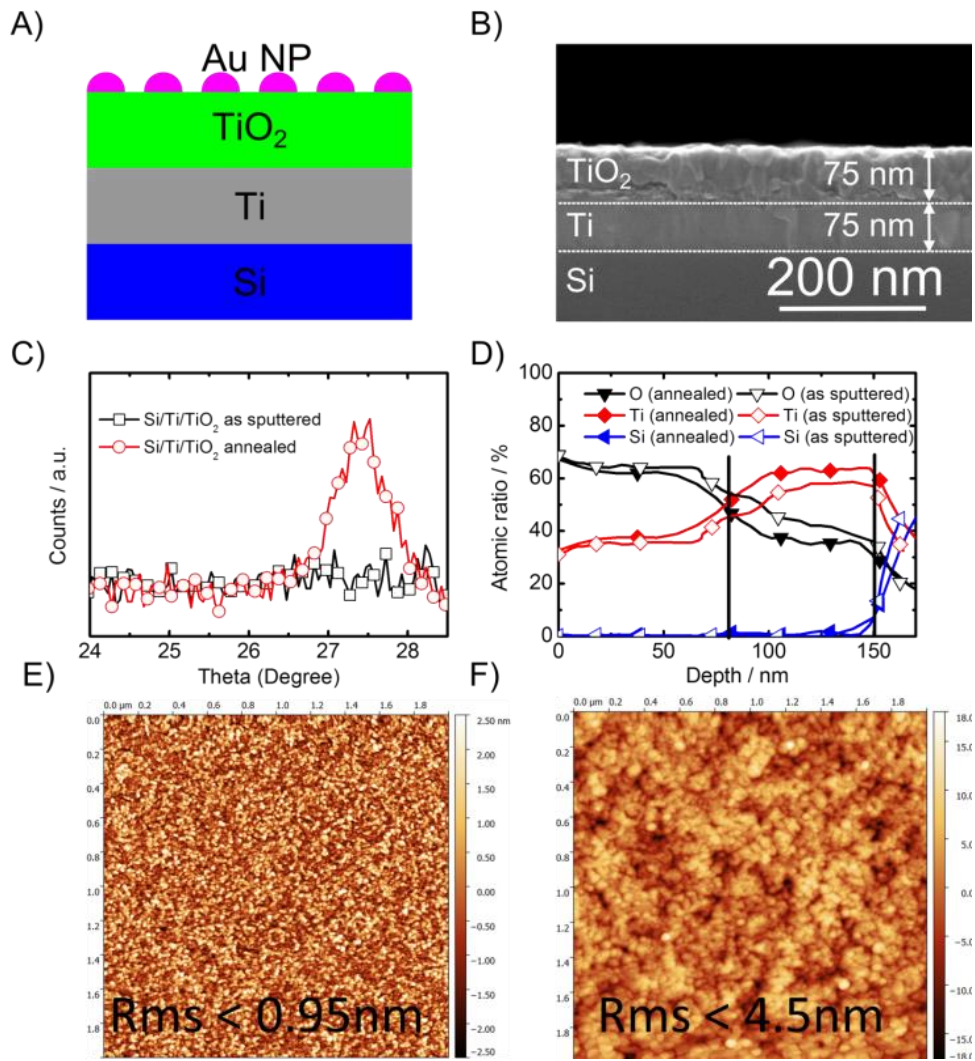

**Figure 2**

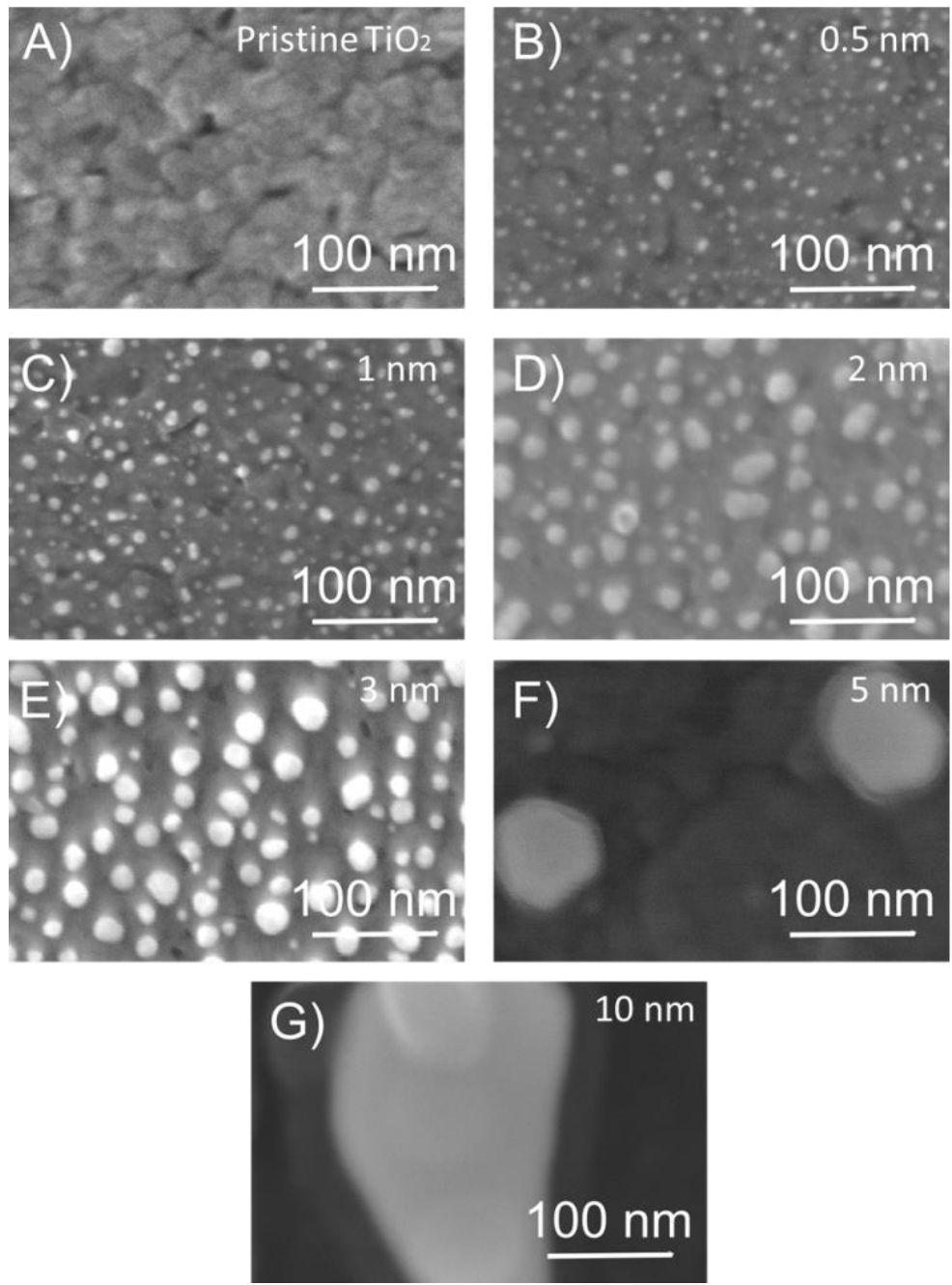

**Figure 3**

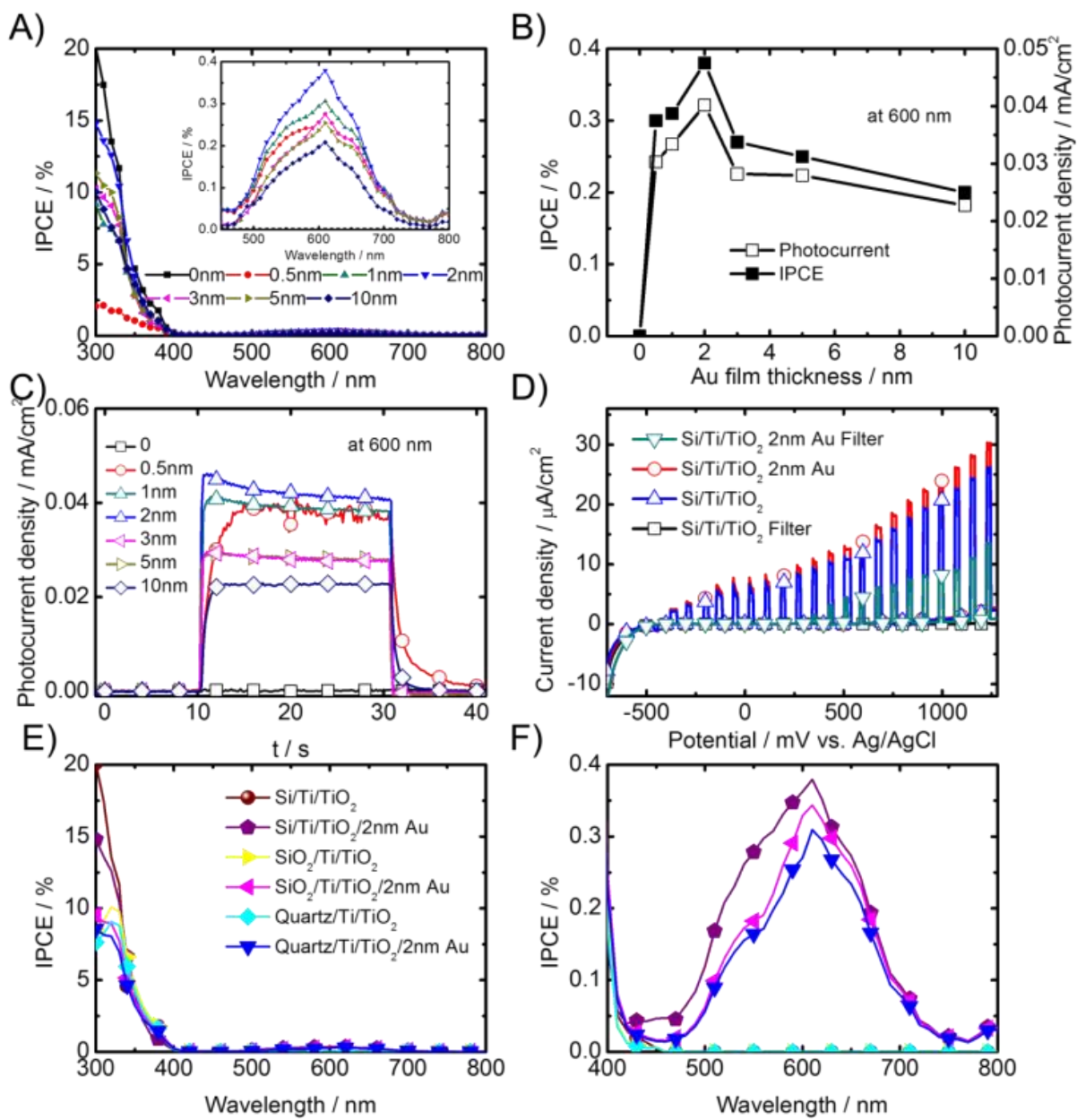

**Table of contents (TOC)**

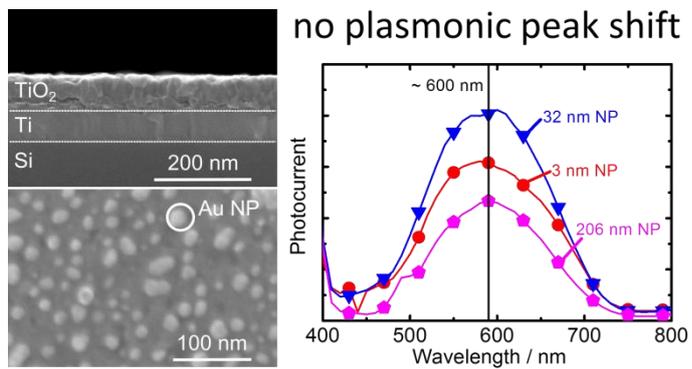